\documentclass[twocolumn]{aastex63}
\usepackage{diagbox}
\usepackage{multirow}
\usepackage{bm}

\received{----}
\revised{----}
\accepted{----}

\begin{document}

\title{On the Binary-Neutron-Star Post-Merger Magnetar Origin of XRT 210423}

\author{Shunke Ai}
\affiliation{Department of Physics and Astronomy, University of Nevada Las Vegas, Las Vegas, NV 89154, USA \\ ais1@unlv.nevada.edu, zhang@physics.unlv.edu}

\author{Bing Zhang}
\affiliation{Department of Physics and Astronomy, University of Nevada Las Vegas, Las Vegas, NV 89154, USA \\ ais1@unlv.nevada.edu, zhang@physics.unlv.edu}

\begin{abstract}
XRT 201423 is an X-ray transient with a nearly flat plateau lasting 4.1 ks followed by a steep decay. This feature indicates that it might come from a magnetar formed through a binary neutron star merger, similar to CDF-S XT2 and as predicted as a type of electromagnetic counterpart of binary neutron star mergers. We test the compliance of the data with this model and use the observed duration and flux of the X-ray signal as well as upper limits of optical emission to pose constraints on the parameters of the underlying putative magnetar. Both the free-zone and trapped-zone geometric configurations are considered. We find that the data are generally consistent with such a model. The surface dipolar magnetic field and the ellipticity of the magnetar should satisfy $B_p < 7\times 10^{14}{\rm G}$ ($B_p < 4.9 \times 10^{14}{\rm G}$) and $\epsilon < 1.5 \times 10^{-3}$ ($\epsilon < 1.1 \times 10^{-3}$) under free zone (trapped zone) configurations, respectively. An upper limit on the distance (e.g. $z < 0.55$ with $\eta_x = 10^{-4}$ or $z < 3.5$ with $\eta_x = 10^{-2}$) can be derived from the X-ray data which depends on the X-ray dissipation efficiency $\eta_x$ of the spin-down luminosity. The non-detection of an optical counterpart places a conservative lower limit on the distance of the source, i.e. $z > 0.045$ regardless of the geometric configuration.
\end{abstract}

\section{Introduction}
On 2021 April 23, Chandra X-ray Observatory serendipitously detected a fast X-ray transient (XRT) in a calibration observation of Abell 1795, which is named as XRT 210423 \citep{lin21}. The rising of the signal is fast, within a few tens of seconds. Then, it maintained at a roughly constant flux (a lightcurve plateau) which lasted for $\sim 4.1{\rm ks}$ and followed by a steep power-law decay in flux as $\propto t^{-3.6}$. Optical observations have been made with Zwicky Transient Facility (ZTF), Xinglong 2.16m Telescope, Large Binocular Telescope and 200-inch Hale Telescope at Palomar Observatory, before and after the X-ray signal was triggered. However, no counterpart was found \citep{andreoni21,xin21,andreoni21b,rossi21}. This XRT is quite similar to another XRT discovered from Chandra Deep Field South, CDF-S XT2 \citep{xue19}. The lightcurves of both transients are consistent with spindown of a rapidly spinning magnetar. Such a short-GRB-less XRT has been predicted to be an electromagnetic counterpart of binary neutron star (BNS) merger gravitational wave events \citep{zhang13}. In the case of CDF-S XT2, a host galaxy at $z = 0.738$ was identified \citep{guo13,santini15,zheng17,xue19}. The source of CDF-S XT2 has a moderate offset from the galaxy center, which is consistent with the location of short GRB sources, and hence, a BNS merger origin \citep{xue19}. The estimated event rate density is also consistent with that of BNS mergers derived from the gravitational wave data \citep{abbott19,xue19}, lending further support to the scenario. Another XRT discovered from Chandra Deep Field South, CDF-S XT1 \citep{bauer17} may be also understood within the same framework if one considers more complicated viewing geometry of the system \citep{sun17,sun19}, even though other possibilities, e.g. a short GRB seen off axis, a distant low-luminosity GRB, a tidal disruption event involving an intermediate mass black hole and a white dwarf, remain possible \citep{bauer17,sarin21}.

The BNS post-merger magnetar scenario for XRT 210423 is of a special interest, as it is closely related to GW-led multi-messenger astronomy, as exemplified by the detection of GW170817/GRB 170817A association \citep{abbott17a,abbott17b,goldstein17,zhang18}. In particular, the existence of a massive NS following a BNS merger would require a stiff NS equation of state (EoS), which is currently not identified. Short GRB observations already indicate possible late-time central engine activities that might require a massive neutron star post-merger product \citep{dai06,fanxu06,gaofan06,metzger08}. In particular, the existence of the so-called internal X-ray plateau following tens of percent of short GRBs \citep{rowlinson10,rowlinson13,lu15}  suggest that a good fraction of BNS mergers leave behind supramassive or even stable neutron stars. Under this interpretation, the NS EoS can be constrained to have a maximum mass as $M_{\rm TOV} \sim 2.3 M_\odot$ \citep[e.g.][]{lasky14,gao16,lu15}. If this is the case, a good fraction of BNSs would make massive neutron stars \citep[e.g.][]{dai06,zhang13,giacomazzo13,radice18}. Most of them might be viewed at large viewing angles so that a short GRB might not be observable. In this case, a magnetar-powered X-ray EM counterpart might be associated with the GW event \citep{zhang13,metzger14,siegel16a,siegel16b,sun17,sun19}. The optical light due to r-process and radioactive decay accompanied with the merger event (the so-called kilonova, \citealt{lipaczyski98,metzger10}) would be also enhanced by energy injection of the post-merger magnetar engine \citep{yu13,metzger14,wollaeger19}. Such a magnetar-powered optical transient was also called a mergernova \citep{yu13}. Evidence of mergernova was collected in several short GRBs showing a shallow decay or plateau feature in the X-ray afterglow \citep[e.g.][]{fan13,gao17b}, and it has been argued that the kilonova associated with GW170817/GRB 170817A was also consistent with an engine-powered mergernova \citep{yu18,li18}. 

In this {\em Letter}, we discuss the compliance of BNS post-merger magnetar model for XRT 210423 using the available X-ray and optical data. In section \ref{sec:Xray}, we constrain the magnetar properties based on the X-ray data. In section \ref{sec:mergernova}, we constrain the distance of the source based on the optical upper limits. Conclusions are presented in section \ref{sec:conclusion}. We take the convention $Q_{m} = Q/10^m$ in cgs units, except $M_{\rm ej,m} = M_{\rm ej}/(10^m M_{\odot})$.

\section{X-ray emission}
\label{sec:Xray}
For a BNS post-merger magnetar, X-ray emission can in principle originate from one of three zones. From small to large of the viewing angle with respect to the jet axis, they are \citep{sun17}: 1. the jet zone, where a bright short GRB is also detected; 2. the free zone, where a short GRB is not detected but X-rays can escape freely shortly after the merger; and 3. the trappd zone, where X-rays are initially trapped in the kilonova ejecta and only become observable when the emission becomes transparent. Notice that there could be weak jet emission in the free zone. For example, GRB 170817A would become non-detectable by Fermi GBM if the distance is greater than 65 Mpc \citep{zhang18}, or $z>0.015$. Yet X-rays can escape freely since a weak GRB already cleared the path. Since the viewing angle of GRB 170817A could be as large as $\sim 30^{\rm o}$ \citep[e.g.][]{troja19,song19}, the X-ray free zone could occupy a significant solid angle for BNS mergers at cosmological distances. Assume that XRT 210423 is associated with a BNS merger event. Since no short GRB was detected, only the last two possibilities are relevant.

The decay index of XRT 201423, $-3.6$ \citep{lin21}, is steeper than $-2$ or $-1$, the predicted magnetar spindown index due to magnetic dipole radiation or gravitational wave radiation, respectively \citep{zhangmeszaros01,lasky16}. Such a feature is often observed as the internal plateau in short GRB afterglows \citep{rowlinson10,rowlinson13,lu15}, which may be interpreted as the collpase of the supramassive neutron star to a black hole after it spins down \citep{zhang14,lasky14,gao16,breurezzolla16,ai20}. In this case, the break time at the end of plateau is a lower limit of the spindown timescale, but since the SMNS is significantly spun down only at around the spindown timescale, this timescale is not too much smaller than the spindown timescale \citep{ravi14}. In the following, we use these arguments to constrain magnetar parameters for both the free zone and trapped zone geometries. 

\subsection{Free zone}
\label{sec:freezone}
In the free zone, spin-down powered X-ray emission can escape freely. Hence, the onset of the X-ray emission roughly corresponds to the merger time. 

A newly born NS can release its rotational energy through both electromagnetic (EM) waves and gravitational waves (GW). The general expression for the spin-down luminosity could be expressed as \citep{shapiro83,zhangmeszaros01}
\begin{eqnarray}
L_{\rm sd} &=& L_{\rm sd,EM} + L_{\rm sd,GW} \nonumber \\
&=& {B_p^2 R_s^6 \Omega^4 \over 6c^3} + {32GI^2\epsilon^2 \Omega^6 \over 5c^5}
\end{eqnarray}
where $L_{\rm sd,EM}$ and $L_{\rm sd,GW}$ represent the spin-down luminosity introduced by magnetic dipole radiation and gravitational wave radiation, respectively, $B_p$ is the strength of dipolar magnetic field at the polar cap on the NS surface, $\epsilon$ is the ellipticity, a parameter to describe the deformation of the NS, $R_s$ is the NS radius, $\Omega = 2\pi/P$ is the angular velocity of the rotating magnetar, and $P$ is the spin period. Adopting the fiducial values $R_s = 10^{6}{\rm cm}$ and $P = 1{\rm ms}$ for a new-born NS, the spin-down luminosity due to dipole radiation can be estimated as
\begin{eqnarray}
L_{\rm sd,EM} = 10^{49} R_{\rm s,6}^{6} B_{\rm p,15}^2 P_{-3}^{-4} ~{\rm erg~s^{-1}}.
\label{eq:Lsd}
\end{eqnarray}
The corresponding spin-down timescale is
\begin{eqnarray}
t_{\rm sd,EM} = {E_{\rm rot} \over L_{\rm sd,EM}} = 2.0\times 10^3 I_{\rm 45} R_{\rm s,6}^{-6} B_{\rm p,15}^{-2} P_{\rm i,-3}^2 ~{\rm s},
\end{eqnarray}
where  $I$ is the momentum of inertia, and $E_{\rm rot} = (1/2)I\Omega_0^2 \approx 2\times 10^{52} I_{45}P_{\rm i,-3}^{-2} {\rm erg}$ is the initial total rotational energy of the magnetar. Similar, the GW spin-down timescale is
\begin{eqnarray}
t_{\rm sd,GW} = 9.1 \times 10^3 I_{45}^{-1}P_{i,-3}^{4}\epsilon_{-3}^{-2}~{\rm s}
\end{eqnarray}

For XRT 210423, the duration of the X-ray plateau is $T_X = 4.1\times 10^3 {\rm s}$. Since there is no significant spin-down during the plateau stage, where $P = P_i$ can be assumed, one can deduce that both of the two spin-down timescales should be greater than the collapsing timescale, thus placing upper limits on both $B_p$ ($t_{\rm sd,EM}< T_X$) and $\epsilon$ ($t_{\rm sd,GW}<T_X$), i.e.
\begin{eqnarray}
B_p < 7 \times 10^{14} R_{\rm s,6}^{-3} P_{\rm i,-3} I_{45}^{1/2} ~{\rm G},
\label{eq:Bup}
\end{eqnarray}
and
\begin{eqnarray}
\epsilon < 1.5 \times 10^{-3} I_{45}^{-1/2}P_{\rm i,-3}^2.
\end{eqnarray}

Theoretically, only a fraction of spin-down luminosity in the EM channel can be dissipated into X-ray emission. We can therefore write
\begin{eqnarray}
L_X = \eta_x L_{\rm sd,EM} = 10^{45} \eta_{x,-4} R_{\rm s,6}^{6} B_{\rm p,15}^2 P_{-3}^{-4} ~{\rm erg~s^{-1}}.
\label{eq:theory}
\end{eqnarray}
Observationally, the flux of the X-ray plateau is determined by both the X-ray Luminosity ($L_X$) and the distance of the source, which can be expressed as
\begin{eqnarray}
L_X = 4\pi d_L^2 F_X = 5.0\times 10^{44}d_{\rm L,28}^2~{\rm erg~s^{-1}},
\label{eq:observation}
\end{eqnarray}
where $F_X = 4\times 10^{-13}~{\rm erg~s^{-1}~cm^{-2}}$ is the observed X-ray flux \citep{lin21}, and $d_{\rm L,28}$ is the unknown luminosity distance of the source. Combining Equation \ref{eq:theory} and Equation \ref{eq:observation}, the magnetic field strength can be calculated as
\begin{eqnarray}
B_p = 7.0 \times 10^{14}\eta_{x,-4}^{-1/2} R_{\rm s,6}^{-3} P_{-3}^2 d_{\rm L,28} ~{\rm G}.
\label{eq:B}
\end{eqnarray}
Comparing Equation \ref{eq:Bup} with Equation \ref{eq:B}, once can estimate the upper limit on the distance
\begin{eqnarray}
d_L < 3.2 \eta_{x,-4}^{1/2}P_{i,-3}^{-1} I_{45}^{1/2}~{\rm Gpc},
\end{eqnarray}
or $z < 0.55$ for the adopted parameters. We note that this result sensitively depends on the uncertain parameter $\eta_x$. For example, if $\eta_x = 10^{-2}$, the redshift upper limit becomes $z<3.5$, which is essentially unconstrained. 

\subsection{Trapped zone}
\label{sec:trappedzone}
In the trapped zone, the spin-down-powered X-ray emission was initially trapped by the optically thick ejecta. The optical depth of the ejecta could be estimated as
\begin{eqnarray}
\tau = \kappa (M_{\rm ej}/V)\Delta_{\rm ej},
\label{eq:tau}
\end{eqnarray}
where $\kappa$ is the opacity \citep{kasenbildsten10,kotera13}, $M_{\rm ej}$ is the ejecta mass, $V = 4\pi R^2 \Delta_{\rm ej}$ is the volume of the ejecta in the lab frame with $\Delta_{\rm ej}$ and $R$ being the thickness of the shell and the distance from the shell to the central magnetar, respectively. At early times, the volume of the ejecta is small with $\tau \gg 1$. The spin-down energy would be injected into the ejecta and be converted to  internal and kinetic energy. 

Here we estimate the time when the ejecta becomes optically thin analytically. Since the ejeca is not exactly isotropic, one could define different effective ejecta masses ($M_{\rm ej,eff}$) for different viewing angles (which is defined as the isotropic ejecta mass if the ejecta mass per solid angle is the same as the one along the line of sight) and the total ejecta mass $M_{\rm ej}$ (which is defined as the angle-integrated ejecta mass) \citep[e.g.][]{sun19}. The latter is relevant to calculating the brightness of the kilonova/mergernova while the former is relevant to calculating the transparent time of non-thermal X-rays along the line of sight. From Equation \ref{eq:tau}, we find the radius for $\tau \sim 1$ is roughly\footnote{The mergernova ejecta is assumed to be homologous with a distribution of the expansion speed. The thickness of the shell therefore continuously increases with time. If the maximum and minimum speeds differ significantly by over an order of magnitude, the outer front of the shell continuously expands while the inner bound of the shell advances slowly. The thickness of the shell is approximately the radius of the shell front, i.e. $\Delta_{\rm ej} \sim R$, so that the ejecta is essentially a sphere. In the thin-shell approximation (assuming a uniform speed of the ejecta), one would have $\Delta_{\rm ej} \ll R$. The spherical-ejecta treatment would make the estimation of $R_{\tau}$ greater by a factor $\sqrt{3}$.}
\begin{eqnarray}
R_{\tau} \sim 4.0 \times 10^{14} \kappa^{1/2} M_{\rm ej,eff,-3}^{1/2}~{\rm cm}.
\label{eq:Rtau}
\end{eqnarray}
Assume that a good fraction $\xi_k$ of EM spin-down luminosity can be converted to the kinetic energy of the ejecta and ignore the radioactive power and  represent the Lorentz factor of the ejecta as $\Gamma$. For relatively low effective ejecta mass ($M_{\rm ej,eff} \lesssim 10^{-3}M_{\odot}$), we can take the relativistic approximation ($\Gamma \gg 1$).
 
From energy conservation, one has
\begin{eqnarray}
\xi_k L_{\rm sd,EM}t = (\Gamma-1)M_{\rm ej,eff}c^2.
\end{eqnarray}
The radius of the ejecta at $\tau = 1$ can be also expressed as
\begin{eqnarray}
R_{\tau} &\sim& 2\Gamma_{\tau}^2 c t_{\tau} \nonumber \\
&\sim& {2\xi_k^2 L_{\rm sd,EM}^2 t_{\tau}^3 \over M_{\rm ej,eff}^2 c^3},
\label{eq:Rtau1}
\end{eqnarray}
where $\Gamma_{\tau} \sim \xi_k L_{\rm sd,EM}t_{\tau}/(M_{\rm ej,eff} c^2)$ with $t_{\tau}$ representing the time when $\tau = 1$. From Equation \ref{eq:Rtau} and Equation \ref{eq:Rtau1}, the optically thin timescales could be written as
\begin{eqnarray}
t_{\tau} \sim 1.3 \times 10^4 \xi_k^{-2/3} M_{\rm ej,eff,-3}^{5/6} R_{\rm s,6}^{-4} B_{\rm p,14}^{-4/3} P_{i,-3}^{8/3}~{\rm s}.
\label{eq:ttau}
\end{eqnarray}

For XRT 210423, Since $L_X \propto t^0$ at the plateau, we should have both $t_{\rm sd,EM}$ and $t_{\rm sd,GW}$ greater than $t_{\tau} + T_X$.
\begin{itemize}
    \item If $t_{\tau} \ll T_X$, which requires a very low effective ejecta mass (i.e. the line of sight enters the trapped zone where the ejecta density and effective masses are still low), the results are similar to the free zone case. The constraints on $B_p$ and $\epsilon$ are the same as that in section \ref{sec:freezone}.
    \item If $t_{\tau} \gg T_X$ (i.e. the line of sight is deep in the trapped zone), the constraints are very different from the free zone case. First of all, one can constrain $B_p$ directly from the  $t_{\tau} \gg T_X$ relation as
    \begin{eqnarray}
    B_p \ll 2.4 \times 10^{14} \xi_k^{1/2} M_{\rm ej,eff,-3}^{5/8} R_{s,6}^{-3}P_{i,-3}^2 {\rm G},
    \label{eq:ttaugreat}
    \end{eqnarray}
    which means that the merger product is likely not a magnetar. 
    From $t_{\rm sd,EM}>t_{\tau}$, one cannot find further constraint, unless the effective mass is much higher than $10^{-3}M_{\odot}$. From $t_{\rm sd,GW} > t_{\tau}$ and Equation (\ref{eq:ttaugreat}), one can constrain the ellipticity to
    \begin{eqnarray}
    \epsilon \ll 1.5 \times 10^{-3} \xi_k^{2/3} I_{45}^{-1/2}P_{i,-3}^2,
    \end{eqnarray}
    which is similar to the constraint in the free-zone case (Section \ref{sec:freezone}).
    Notice that the $t_{\tau} \gg T_X$ condition requires a contrived condition to allow the difference between two large numbers being a small number. 
    \item The $t_{\tau} \sim T_X$ situation is more natural to avoid the fine-tuning problem in the $t_{\tau} \gg T_X$ case. In this case, the constraints become $t_{\rm sd} \gtrsim 2T_X$. The upper limits on $B_p$ and $\epsilon$ are
    \begin{eqnarray}
    B_p < 4.9 \times 10^{14} I_{45}^{1/2}R_{s,6}^{-3}P_{i,-3},
    \label{eq:B_trap}
    \end{eqnarray}
    and
    \begin{eqnarray}
    \epsilon < 1.1 \times 10^{-3} I_{45}^{-1/2} P_{i,-3}^2,
    \end{eqnarray}
    respectively. 
    Substituting Equation (\ref{eq:B}) into Equation (\ref{eq:B_trap}), one could constrain the distance of the source as
    \begin{eqnarray}
    d_L < 2.3 \eta_{x,-4}^{1/2} I_{45}^{1/2}P_{i,-3}^{-1}~{\rm Gpc},
    \end{eqnarray}
    or $z<0.4$ for $\eta_x=10^{-4}$ and $z<2.7$ for $\eta_x=10^{-2}$. 
\end{itemize}

The expression of $t_\tau$ above (Equation \ref{eq:ttau}) is only valid when the energy injection from the central magnetar is significant. For the low $B_p$ case, the expansion of the ejecta will be dominated by radioactive power and the transparent time could be estimated as \citep{metzger10}
\begin{eqnarray}
t_{\tau}=1.25\times 10^{5}&&\left(\frac{v_{\rm ej}}{0.1c}\right)^{-1/2}\left(\frac{M_{\rm ej,eff}}{10^{-2}M_{\odot}}\right)^{1/2} \nonumber\\
&&\left(\frac{\kappa}{1~{\rm cm^2~g^{-1}}}\right)^{1/2}~{\rm s},
\end{eqnarray}
In this case, in order to allow $t_{\tau} \lesssim T_X$ to avoid the fine-tuning problem, $M_{\rm ej,eff} \lesssim 10^{-4}M_{\odot}$ is required.

\section{Mergernova}
\label{sec:mergernova}
Optical observations in multiple bands have been conducted near the emerging time of XRT 210423 by ZTF and other telescopes \citep{andreoni21,xin21,rossi21,andreoni21b}. However, no signal exceeding the limiting magnitude was detected. In this section, we investigate how the non-detection poses constraints on the magnetar origin of XRT 210423.

To calculate the lightcurve of mergernova, one should fully consider the dynamical evolution of the ejecta. We follow the analytical model proposed by \cite{yu13} with some refined treatments. From energy conservation, we have
\begin{eqnarray}
{dE_{\rm ej} \over dt} = L_{\rm sd,EM} + L_{\rm ra} - L_e,
\label{eq:dEej}
\end{eqnarray}
where $L_{\rm ra}$ is the radioactive luminosity due to r-process nucleosynthesis and $\beta$-decay, $L_e$ is the bolometric luminosity of emission from the ejecta. For a hydrodynamic ejecta, the energy density should be equal to the 00 component of the energy-momentum tensor, which reads as
\begin{eqnarray}
T_{00} = \Gamma^2 (\rho^{\prime} c^2+e^{\prime}+p^{\prime})-p^{\prime},
\end{eqnarray}
where $\Gamma$ is the Lorentz factor of the ejecta, $\rho^{\prime}$, $e^{\prime}$ and $p^{\prime}$ are the mass density, thermal energy density and pressure in rest frame of the ejecta (hereafter the primed parameters are all defined in the comoving frame of the ejecta).
Consider that $p^{\prime} = (\hat{\gamma}-1)e^{\prime}$, where $\hat{\gamma}$ is the adiabatic index. The total energy of the ejecta may be also expressed as\footnote{In \cite{yu13}, the total energy of the ejecta was approximately written as $E_{\rm ej} = (\Gamma - 1)M_{\rm ej} c^2 + \Gamma E_{\rm int}^{\prime}$. A more accurate expression should also include the equation of state of the ejecta.}
\begin{eqnarray}
E_{\rm ej} &=& T_{00}V \nonumber \\
&=&(\Gamma^2 (\rho^{\prime} c^2+e^{\prime}+p^{\prime})-p^{\prime}){V' \over \Gamma}\nonumber \\
&=&\Gamma M_{ej}c^2 + \left[\Gamma\hat{\gamma} - {(\hat{\gamma}-1) \over \Gamma}\right]e^{\prime}V' \nonumber \\
&=&\Gamma M_{\rm ej} c^2 + \Gamma_{\rm eff} E_{\rm int}^{\prime},
\label{eq:Eej}
\end{eqnarray}
where the effective Lorentz factor is defined as
\begin{eqnarray}
\Gamma_{\rm eff}=\Gamma \hat{\gamma} - {\hat{\gamma}-1 \over \Gamma}.
\end{eqnarray}
For a GW170817-like event, the ejecta is relatively massive ($>10^{-2}M_{\odot}$) with a non-relativistic bulk motion ($v_{\rm ej} < 0.3c$). However, under the thick-shell assumption ($\Delta_{\rm ej} \sim R$), the internal energy is dominated by radiation, so that a relativistic gas is considered with $\hat{\gamma} = 4/3$ in the rest of discussion. Calculating the time derivative of Equation \ref{eq:dEej} and substituting it into Equation \ref{eq:Eej}, one can then obtain the evolution of Lorentz factor 
\begin{eqnarray}
{d\Gamma \over dt} = {L_{\rm sd,EM} + L_{\rm ra} - L_e - \Gamma_{\rm eff} {\cal D}(dE_{\rm int}^{\prime}/dt^{\prime}) \over M_{\rm ej}c^2 + (\hat{\gamma} + (\hat{\gamma}-1)/\Gamma^2)E_{\rm int}^{\prime}}
\label{eq:dGamma}
\end{eqnarray}
where ${\cal D} = 1/[\Gamma(1-\beta)]$ is the Doppler factor with $\beta = \sqrt{1-\Gamma^{-2}}$. Consider that a fraction $\xi$ of the spin-down luminosity could be converted to internal energy of the ejecta. Also consider radioactive heating as well as cooling via bolometric emission and $p dV$ work, the evolution of internal energy could be written as
\begin{eqnarray}
{dE_{\rm int}^{\prime} \over dt^{\prime}} = \xi L_{\rm sd}^{\prime} + L_{\rm ra}^{\prime} - L_e^{\prime} - p^{\prime} {dV^{\prime} \over dt^{\prime}}.
\label{eq:dEint}
\end{eqnarray}
where $L^{\prime} = L/{\cal D}^2$. The radioactive luminosity is estimated as \citep{korobkin12}
\begin{eqnarray}
L_{\rm ra}^{\prime} = 4\times 10^{49} M_{\rm ej,-2} \times \left[{1\over 2}-{1 \over \pi} {\rm arctan}({t^{\prime} - t_0^{\prime} \over t_{\sigma}^{\prime}}) \right]^{1.3} {\rm erg~s^{-1}} \nonumber \\
\end{eqnarray}
with $t_0^{\prime} \sim 1.3~{\rm s}$ and $t_{\sigma}^{\prime} = 0.11~{\rm s}$.
For a relativistic, ideal gas, the relation between pressure and internal energy may be generally written as
\begin{eqnarray}
p^{\prime} = {E_{\rm int}^{\prime} \over 3V^{\prime}},
\end{eqnarray}
Consider a homologous explosion with a distribution of ejecta speed with $v_{\rm max} \gg v_{\rm min}$. The ejecta can be treated as a thick shell with a nearly spherical shape. Define an effective radius $R$ and speed $\beta$ of the mergernova for the entire thick shell. To first-order, one may approximate $\beta \sim \beta_{\rm max}$ and $R \sim R_{\rm max}$, where $R_{\rm max} = \beta_{\rm max} c t$ is the radius of the fastest part of the ejecta. 
The comoving volume and radius could be calculated by \citep{yu13}
\begin{eqnarray}
{dV^{\prime} \over dt^{\prime}} = 4\pi R^2 \beta c,
\label{eq:dV}
\end{eqnarray}
and 
\begin{eqnarray}
{dR \over dt} = {\beta c \over 1-\beta}.
\label{eq:dR}
\end{eqnarray}
The bolometric luminosity can be estimated from the internal energy as 
\begin{eqnarray}
L'_e=\left \{ \begin{array}{ll} 
\Gamma E'_{\rm int}c/(\tau R), & ~~\textrm{for $ \tau>1$},\\
\\
\Gamma E'_{\rm int}c/R, & ~~\textrm{for $ \tau \leq 1$}.\\
\end{array} \right.
\end{eqnarray}
Solving the differential equations (\ref{eq:dGamma}), (\ref{eq:dEint}), (\ref{eq:dV}) and (\ref{eq:dR}), one can trace the dynamical evolution of the ejecta.

The effective temperature may be estimated as
\begin{eqnarray}
T^{\prime} = \left({L_e^{\prime} \over 4\pi R^2 \sigma}\right)^{1/4},
\end{eqnarray}
where $\sigma$ is the Stefan-Boltzmann constant. The specific flux of the mergernova is 
\begin{eqnarray}
F_{\nu} = {1 \over 4\pi d_L^2(1+z)}{8\pi^2 {\cal D}^2 R^2 \over h^3 c^2 \nu_0}{(h\nu_0 / D)^4 \over {\rm exp}(h\nu_0/D k T^{\prime})-1}
\end{eqnarray}
with $h$ as the Planck's constant. $\nu_0 = (1+z)\nu$ is the frequency at the source. 

\begin{figure}
\resizebox{90mm}{!}{\includegraphics[]{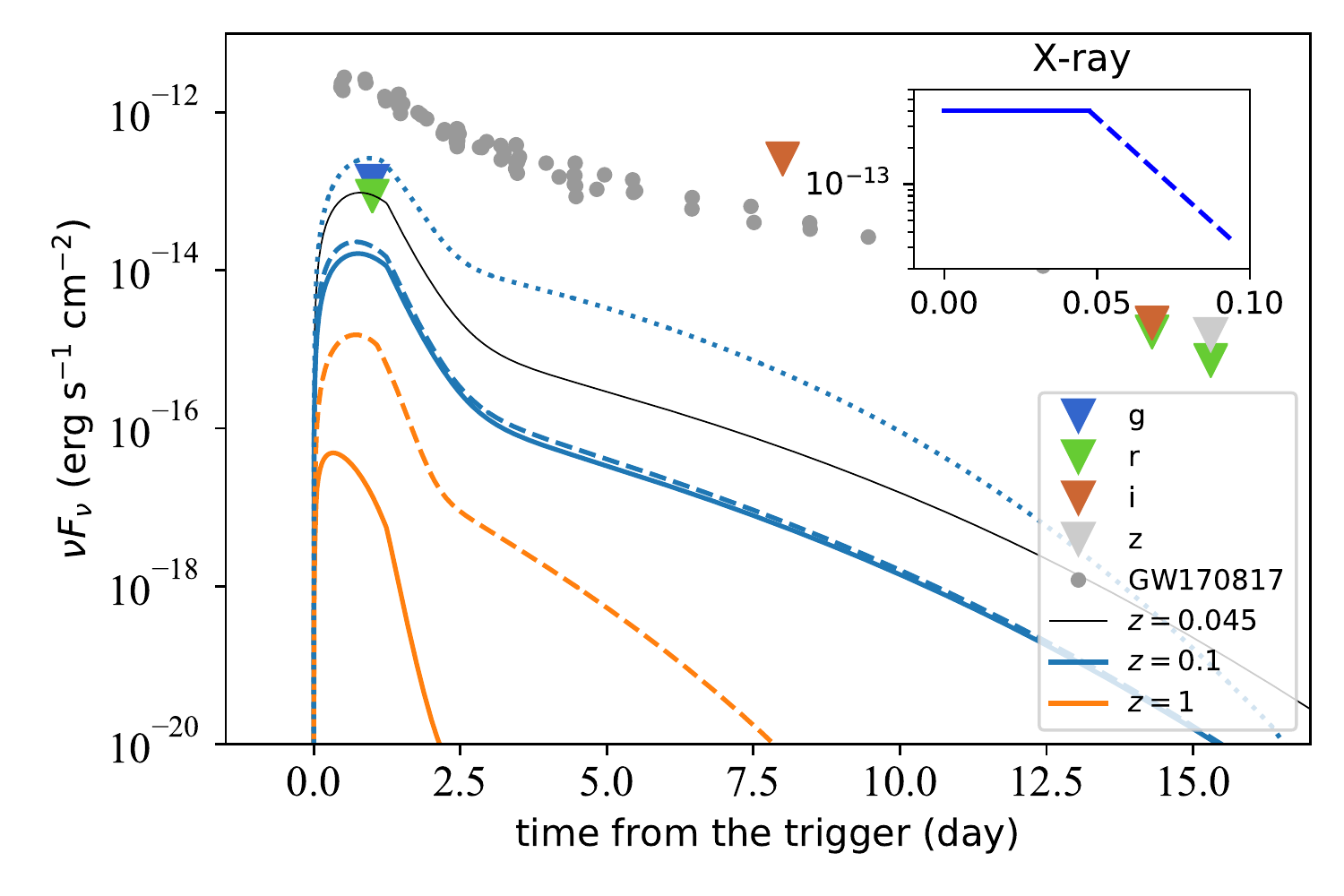}}
\caption{The predicted lightcurves of a putative mergernova associated with XRT 210423, under the free zone assumption. The inverted triangles represent the upper limits obtained from the optical observations \citep{andreoni21,rossi21,andreoni21b}. The solid, dashed and dotted lines stand for the cases of $\eta_x = 1$, $\eta_x = 10^{-2}$ and $\eta_x = 10^{-4}$, respectively. We adopt the magnetar parameters $R_s = 10^6~{\rm cm}$, $P \approx P_i = 1~{\rm ms}$. For the ejecta, two components (blue, red) are considered with $M_{\rm ej,blue} = 0.02 M_{\odot}$, $\kappa_{\rm blue} = 0.5~{\rm cm^2~g^{-1}}$, $\beta_{\rm 0,blue} = 0.3$; $M_{\rm ej,red} = 0.04 M_{\odot}$, $\kappa_{\rm red} = 5~{\rm cm^2~g^{-1}}$, $\beta_{\rm 0,red} = 0.1$. The lightcurve of the optical counterpart of GW170817 is also shown \citep{villar17}. All the theoretical lines and the GW170817 data points are in the r band and $\xi = 0.3$ is assumed.}
\label{fig:mergernova}
\end{figure}

If the X-ray emission is observed from the free zone, the starting time of the X-ray plateau is roughly the merger time of the BNSs. Hence, the optical observations before $t_{X,e}$ would not be relevant, where $t_{X,e}$ is the emerging time of the X-rays. If the X-ray emission is observed from the trapped zone, the merger time should be $t_{X,e} - t_{\tau}$, in which case the optical observations before $t_{X,e}$ would be also relevant. However, as we have discussed in section \ref{sec:trappedzone}, $t_{\tau,X} \lesssim T_X$ would be most naturally expected, which is much smaller than the peak time of a mergernova. Hence, the constraints from optical observations would essentially have no difference between the free zone and the trapped zone cases. Here we only present the results of the former case in Figure \ref{fig:mergernova}.

It is generally believed that the kilonova of GW170817 had two components (``red" and ``blue") \citep{arcavi17,chornock17,cowperthwaite17,drout17,evans17,gao17,kasen17,kilpatrick17,nicholl17,shappee17,smartt17,tanvir17,villar17}. Here we take the two-component model as well as the parameters used to interpret the GW170817 kilonova to calculate our predicted mergernova lightcurves.
As one can see from Figure \ref{fig:mergernova}, the predicted mergernova flux is significantly below the observational upper limits, even if the source is at $z=0.1$. The predicted flux would reach the optical upper limits only if the source is close enough. A conservative redshift lower limit could be posed at $z > 0.045$. This limit is conservative since it depends on the X-ray efficiency. If $\eta_x \ll 1$, with the same distance, the required dipole magnetic field $B_p$ to power the X-ray emission would be also greater, and a brighter mergernova is expected. Therefore, the redshift lower limit is pushed upwards. However, the pair of parameters, $\eta_x$ and distance, are constrained by the spindown timescale. In other words, one cannot impose an extremely small $\eta_x$ to make a mergernova  detectable at a high redshift.

\section{Conclusions and Discussion}
\label{sec:conclusion}
XRT 210423 showed a lightcurve feature consistent with the spindown and probably collapse of a new-born magnetar \citep{lin21}. Prompted by its analogy to CDF-S XT2 \citep{xue19}, which is consistent with a BNS post-merger product as predicted by \cite{zhang13,sun17}, we investigate the compliance of the available X-ray and optical data of XRT 210423 with this model. 

We find that the data are consistent with the model, with the X-rays either observed in the free zone or in the trapped zone with a small effective ejecta mass, so that the true duration of the transient is not much longer than the observed one. We were able to place a list of constraints on the magnetar parameters and the distance of the source according to this model. Specifically, based on X-ray data. we found $B_p < 7.0 \times 10^{14}{\rm G}$, $\epsilon < 1.5 \times 10^{-3}$ for the free zone case, and $B_p < 4.9 \times 10^{14}{\rm G}$, $\epsilon < 1.1\times 10^{-3}$ for the trapped zone case. In principle, an upper limit of the distance can be obtained, but it is highly dependent on the fraction of spin-down luminosity that is dissipated in X-rays. With $\eta_x \sim 10^{-4}$, the source should be approximately at a luminosity distance $d_L < 3.2{\rm Gpc}$ ($z < 0.55$) for the free zone case and $d_L < 2.3{\rm Gpc}$ ($z < 0.4$) for the trapped zone case. However, with $\eta_x \sim 10^{-2}$, the two limits become $z < 3.5$ and $z < 2.7$, respectively, not very constraining. Based on the non-detection of a mergernova in the optical band, the source should be farther away than GW170817, with a conservative redshift lower limit of $z > 0.045$.

The discovery of XRT 210423 \citep{lin21}, together with CDF-S XT2 \citep{xue19}, suggests that X-ray transients with a characteristic magnetar signature are quite common. If they are indeed associated with BNS mergers as suggested \citep{zhang13,sun17}, they will eventually be detected along with gravitational wave signals. The joint observations of wide field X-ray detectors such as Einstein Probe \citep{yuan18}, eROSITA \citep{merloni12}, and THESEUS \citep{amati18} with GW observatories will be fruitful to test and confirm such a connection. If indeed BNS post-merger magnetars are commonly observed in the future, it will have profound implications for understanding BNS merger physics and the neutron star equation of state in connection with the post-merger products \citep[e.g.][]{lasky14,gao16,li16,margalit19,ai20}.  

There are other fast XRTs with plateau emission discovered from archival data \citep[e.g.][]{jonker13,glennie15}. The connection of these events to BNS mergers is not obvious. There are likely more than one progenitor types for plateau XRTs. Future multi-wavelength, multi-messenger observations of XRTs will reveal possible diverse origins of these events.

\section*{Acknowledgements}
We thank He Gao, Yun-Wei Yu, Peter Jonker and a referee for helpful comments. This work is supported by the Top Tier Doctoral Graduate Research Assistantship (TTDGRA) at University of Nevada, Las Vegas.

\end{document}